\documentclass[10pt,twocolumn,letterpaper]{article}

\usepackage{iccv}
\usepackage{times}
\usepackage{epsfig}
\usepackage{graphicx}
\usepackage{amsmath}
\usepackage{amssymb}
\usepackage{graphicx}
\usepackage[accsupp]{axessibility} 
\usepackage{amsmath}
\usepackage{amssymb}
\usepackage{booktabs}
\usepackage[pagebackref=true,breaklinks=true,letterpaper=true,colorlinks,bookmarks=false]{hyperref}

\usepackage[capitalize]{cleveref}
\crefname{section}{Sec.}{Secs.}
\Crefname{section}{Section}{Sections}
\Crefname{table}{Table}{Tables}
\crefname{table}{Tab.}{Tabs.}

\usepackage{graphicx}
\usepackage{enumitem}
\usepackage[]{hyperref}
\usepackage{bbm}
\usepackage{svg}
\usepackage{amsthm}
\theoremstyle{definition}

\usepackage{epsfig,amsmath,amsfonts,epsfig,multirow,makecell,caption,soul,csquotes,color,wrapfig,subcaption,mathtools,bm,spverbatim,booktabs,tcolorbox,diagbox,todonotes}
\usepackage[e]{esvect}

\def\tablename{Table}
\def\figurename{Figure}

\usepackage{caption}
\makeatletter
\newif\if@restonecol
\makeatother

\usepackage[boxed, ruled, vlined, linesnumbered]{algorithm2e}
\SetKwRepeat{Do}{do}{while}

\newenvironment{changemargin}[2]{\begin{list}{}{
	\setlength{\topsep}{0pt}\setlength{\leftmargin}{-8pt}
	\setlength{\rightmargin}{0pt}
	\setlength{\listparindent}{\parindent}
	\setlength{\itemindent}{\parindent}
	\setlength{\parsep}{0pt plus 1pt}
	\addtolength{\leftmargin}{#1}\addtolength{\rightmargin}{#2}
	}\item}
	{\end{list}}

\newcommand{\splus}{{\scaleobj{0.75}{\bm{+}}}}

\newcommand{\sinv}{{\scaleobj{0.85}{\intercal}}}
\newcommand{\mie}{{\em i.e.}\xspace}
\newcommand{\meg}{{\em e.g.}\xspace}
\newcommand{\mcf}{{\em cf.}\xspace}

\usepackage[first=0,last=9]{lcg}
\usepackage{colortbl}
\definecolor{Gray}{gray}{0.8}
\colorlet{Red}{red!10!white}

\usepackage{hyperref}

\newcommand{\msec}[1]{\S\,\ref{#1}}
\newcommand{\mref}[1]{\,\ref{#1}}
\newcommand{\meq}[1]{Eq\,(\ref{#1})}
\newcommand{\mcite}[1]{\,\cite{#1}}

\newcommand{\mct}[1]{({\em #1})}

\usepackage{scalerel}[2016/12/29]

\makeatletter
\providecommand{\leadsfrom}{%
  \mathrel{\mathpalette\reflect@squig\relax}%
}
\newcommand{\reflect@squig}[2]{%
  \reflectbox{$\m@th#1\leadsto$}%
}
\makeatother

\newtcolorbox{mtbox}[1]{left=0.25mm, right=0.25mm, top=0.25mm, bottom=0.25mm, sharp corners, colframe=blue!50!black, boxrule=0.5pt, title={#1}, fonttitle=\bfseries, coltitle=blue!50!black, attach title to upper={\ --\ }}

\def\eqref#1{equation~\ref{#1}}

\def\1{\bm{1}}

\DeclareMathAlphabet{\mathsfit}{\encodingdefault}{\sfdefault}{m}{sl}
\SetMathAlphabet{\mathsfit}{bold}{\encodingdefault}{\sfdefault}{bx}{n}

\def\gA{{\mathcal{A}}}

\def\gD{{\mathcal{D}}}

\def\gL{{\mathcal{L}}}

\def\gN{{\mathcal{N}}}

\def\gS{{\mathcal{S}}}

\def\sE{{\mathbb{E}}}

\usepackage{amssymb}
\usepackage{caption}
\usepackage{multicol}
\usepackage[ruled,linesnumbered]{algorithm2e}
\usepackage[utf8]{inputenc} 
\usepackage[T1]{fontenc}    
\usepackage{hyperref}       
\usepackage{xurl}            
\usepackage{amsmath}
\usepackage{amssymb}
\usepackage{booktabs}       
\usepackage{amsfonts}       
\usepackage{nicefrac}       
\usepackage{microtype}      
\usepackage{xcolor}         
\usepackage[noend]{algpseudocode}

\newtheorem{theorem}{Theorem}[section]

\newtheorem{assumption}[theorem]{Assumption}
\newcommand\Tau{\mathrm{T}}
\newtcolorbox{mybox}{colframe =  blue!40!white, colback = blue!10!white}

\newcommand{\ssl}{SSL\xspace}

\usepackage{stackengine}

\usepackage{multirow}
\newcommand{\minitab}[2][l]{\begin{tabular}{#1}#2\end{tabular}}

\newcommand{\jiang}[1]{#1}

\newcommand{\system}{{\sc Ctrl}\xspace}
\newcommand{\sslbkd}{{\sc SslBkd}\xspace}
\newcommand{\bdenc}{{\sc PoiEnc}\xspace}

\newcommand{\asr}{ASR\xspace}
\newcommand{\er}{ER\xspace}

\newcommand{\acc}{ACC\xspace}

\newcommand{\deleted}[1]{}

\definecolor{revision}{RGB}{0,199,140}

\iccvfinalcopy 

\def\iccvPaperID{9337} 

\ificcvfinal\fi

\begin{document}
\title{An Embarrassingly Simple Backdoor Attack on Self-supervised Learning}



\author{
Changjiang Li$^*$ \quad
Ren Pang$^*$ \quad Zhaohan Xi$^*$ \quad Tianyu Du$^*$\quad
Shouling Ji$^\dag$ \quad
Yuan Yao$^\ddag$ \quad
Ting Wang$^*$
\\
$^*$Pennsylvania State University  \quad
$^\dag$Zhejiang University \quad
$^\ddag$Nanjing University\\
{\tt\small meet.cjli@gmail.com, \{rbp5354, zxx5113, tjd6042\}@psu.edu, sji@zju.edu.cn,} 
{\tt\small y.yao@nju.edu.cn} \\
{\tt\small inbox.ting@gmail.com}
}

\maketitle
\ificcvfinal\thispagestyle{empty}\fi

\begin{abstract}


As a new paradigm in machine learning, self-supervised learning (\ssl) is capable of learning high-quality representations of complex data without relying on labels. In addition to eliminating the need for labeled data, research has found that \ssl improves the adversarial robustness over supervised learning since lacking labels makes it more challenging for adversaries to manipulate model predictions. However, the extent to which this robustness superiority generalizes to other types of attacks remains an open question.



We explore this question in the context of backdoor attacks. Specifically, we design and evaluate \system, an embarrassingly simple yet highly effective self-supervised backdoor attack. By only polluting a tiny fraction of training data ($\le$ 1\%) with indistinguishable poisoning samples, \system causes {\em any} trigger-embedded input to be misclassified to the adversary's designated class with a high probability ($\ge$ 99\%) at inference time. Our findings suggest that \ssl and supervised learning are comparably vulnerable to backdoor attacks.
More importantly, through the lens of \system, we study the inherent vulnerability of \ssl to backdoor attacks. With both empirical and analytical evidence, we reveal that the representation invariance property of \ssl, which benefits adversarial robustness, may also be the very reason making \ssl highly susceptible to backdoor attacks.
Our findings also imply that the existing defenses against supervised backdoor attacks are not easily retrofitted to the unique vulnerability of \ssl.   
Code is available at: \href{https://github.com/meet-cjli/CTRL}{https://github.com/meet-cjli/CTRL}

\end{abstract}

\section{Introduction}
\label{sec:intro}

As a new machine learning paradigm, self-supervised learning (\ssl) has gained tremendous advances recently\mcite{chen:2020:simple, grill:2020:bootstrap, chen:2021:exploring}. Without requiring data labeling or human annotations, \ssl is able to learn high-quality representations of complex data and enable a range of downstream tasks.
In particular, 
contrastive learning, one dominant \ssl approach\mcite{chen:2020:simple, grill:2020:bootstrap, chen:2021:exploring, chen:2020:improved, he:2020:momentum}, performs representation learning by aligning the features\footnote{Below we use the terms ``feature'' and ``representation'' exchangeably.} of the same sample under varying data augmentations (\meg, random cropping) while separating the features of different samples. In many tasks, contrastive learning has attained performance comparable to supervised learning\mcite{grill:2020:bootstrap}. 
Meanwhile, besides obviating the reliance on data labeling, \ssl also benefits the robustness to adversarial perturbation, label corruption, and data distribution shift by making it more challenging for the adversary to influence model predictions directly\mcite{ssl-robustness,zhong2022self}. However, whether this robustness benefit generalizes to other malicious attacks remains an open question.



\begin{figure}[t]
	\centerline{\includegraphics[width=0.75\columnwidth]{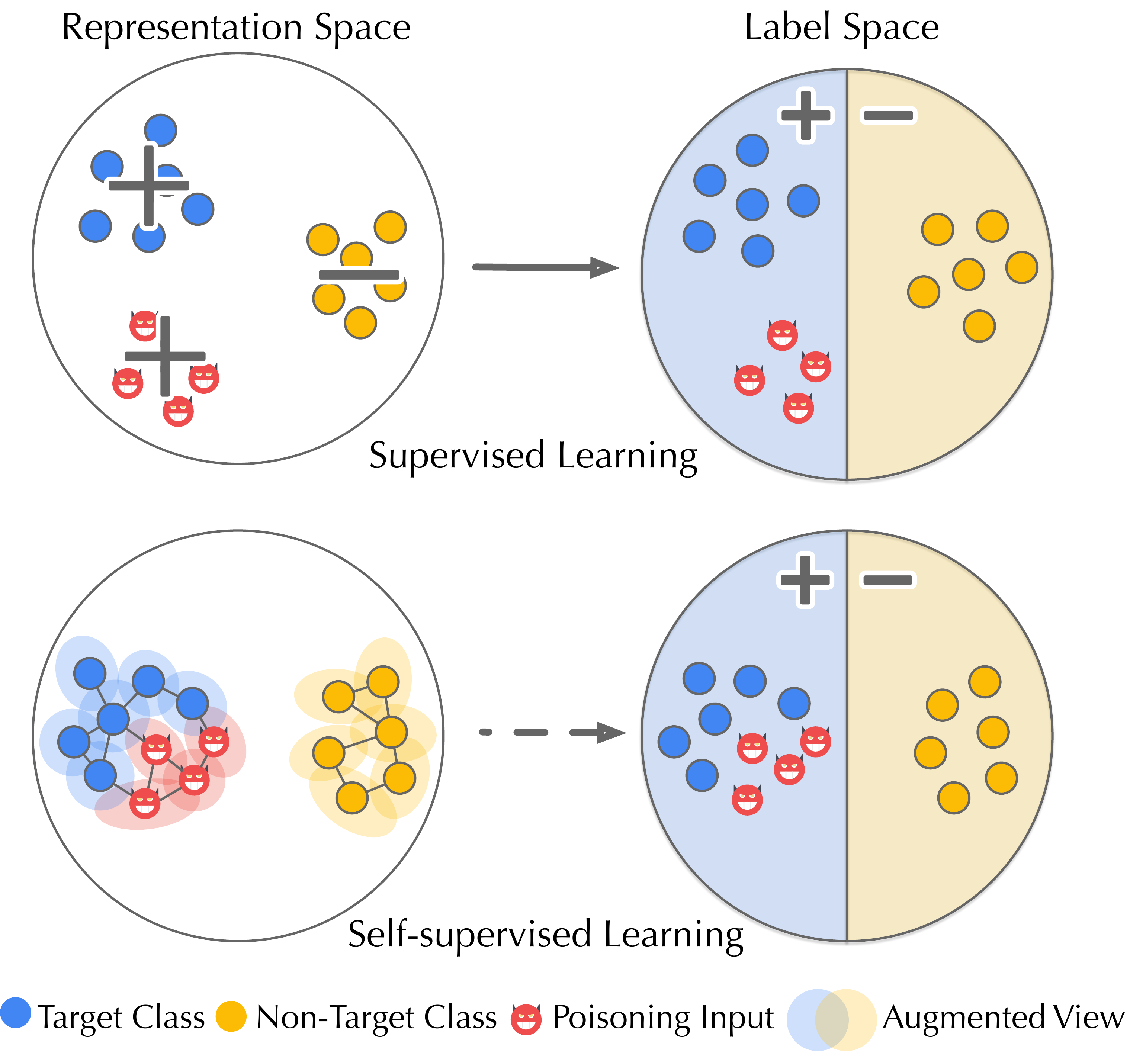}}
	\caption{Comparison of supervised and self-supervised backdoor attacks; self-supervised backdoor attacks influence the label space only indirectly through the representations.}	
	\label{figure:intuition}
\end{figure}

In this work, we explore this question in the context of backdoor attacks, in which the adversary plants ``backdoors'' functions into target models during training and activates such backdoors at inference. 
Recent work has explored new ways to inject backdoors into \ssl-trained models\mcite{saha:2021:backdoor, liu:2022:poisonedencoder,carlini2021poisoning,jia:2021:badencoder}; however, the existing attacks appear to be significantly less effective than their supervised counterparts: they either work for specific, pre-defined inputs only\mcite{jia:2021:badencoder,liu:2022:poisonedencoder} or succeed with a low probability (\meg, $\leq 2$\% on ImageNet-100\mcite{saha:2021:backdoor}). These observations raise a set of intriguing and critical questions: 

\vspace{1pt}
\noindent RQ$_1$ -- {\em Is \ssl comparably vulnerable to backdoor attacks as supervised learning?}

\vspace{1pt}
\noindent RQ$_2$ -- {\em  If so, what makes it highly vulnerable?}

\vspace{1pt}
\noindent RQ$_3$ -- {\em What are the implications of this vulnerability?} 

\vspace{3pt}
{\bf Our Work.} This work represents a solid step toward answering these questions. 

\vspace{1pt}
RA$_1$ -- We present \system\footnote{\system: \underline{C}ontrastive \underline{TR}ojan \underline{L}earning.}, a simple yet highly effective self-supervised backdoor attack. Compared with the existing attacks, \mct{i} \system assumes that the adversary is able to pollute a tiny fraction of training data yet without any control of the training process; \mct{ii} it defines the ``trigger'' as an augmentation-insensitive perturbation in the spectral space of inputs and generates poisoning data indistinguishable from clean data; \mct{iii} it aims to force all trigger-embedded inputs to be misclassified to the adversary's designated class at inference. With evaluation on benchmark models and datasets, we show that \ssl is also highly vulnerable to backdoor attacks. For instance, by poisoning $\leq 1\%$ of the training data, \system achieves $\geq 99\%$ attack success rate on CIFAR-10. This level of vulnerability is comparable to what are observed in supervised backdoor attacks. 

\vspace{1pt}
RA$_2$ -- Through the lens of \system, we study the inherent vulnerability of \ssl. Intuitively, \system exploits data augmentation and contrastive loss, two essential ingredients of \ssl \mcite{chen:2020:simple, grill:2020:bootstrap}, which together entail the representation invariance property: different augmented views of the same input share similar representations. 
Given the overlap between the augmented views of trigger-embedded and target-class inputs, enforcing representation invariance naturally entangles them in the feature space, as illustrated in Figure\mref{figure:intuition}, incurring the risk of backdoor attacks. This mechanism fundamentally differs from supervised backdoor attacks\mcite{Wang:2019:sp, latent-backdoor, imc}, which directly associate the trigger pattern with the target class in the label space, while the representations of trigger-embedded and target-class inputs are not necessarily aligned\mcite{tact}.

\vspace{1pt}
RA$_3$ -- Moreover, we discuss the challenges to defending against self-supervised backdoor attacks. We find that existing defenses against supervised backdoor attacks are not easily retrofitted to the unique vulnerability of \ssl. For instance, {{\sc SCAn}}\mcite{tact}, a state-of-the-art defense, detects trigger-embedded inputs based on the statistical properties of their representations; however, it is ineffective against \system, due to the inherent entanglement between the representations of trigger-embedded and target-class inputs. 


\vspace{3pt}
{\bf Our Contributions.} \jiang{This work establishes a strong baseline for comprehending the inherent vulnerability of SSL to backdoor attacks. By employing innovative techniques and insights, our study contributes to the field in the following ways. }

\vspace{1pt}
We present \system, a simple yet effective self-supervised backdoor attack, which greatly reduces the gap between the attack effectiveness of supervised and self-supervised backdoor attacks. Leveraging \system, we show that \ssl is highly susceptible to backdoor attacks. Our findings imply that the benefit of \ssl for adversarial robustness superiority may not generalize to trojan attacks.

\vspace{1pt}
With both empirical and analytical evidence, we reveal that \mct{i} self-supervised backdoor attacks may function by entangling the representations of trigger-embedded and target-class inputs; \mct{ii} the representation invariance property of \ssl, which benefits adversarial robustness, may also account for the vulnerability of \ssl to backdoor attacks.


\vspace{1pt}
We evaluate \ssl on some existing defenses and point out several promising directions for further research.

\section{Related Work}
\label{sec:literature}

\deleted{In addition to the aforementioned related work, we further survey the literature most relevant to this work. }


\subsection{Self-supervised Learning} Recent years are witnessing the striding advances of self-supervised learning (\ssl)\mcite{chen:2020:simple,moco,grill:2020:bootstrap, chen:2021:exploring}. Using the supervisory signals from the data itself, \ssl trains a high-quality encoder $f$ that extracts high-quality representations of given data, which can then be integrated with a downstream classifier $g$ and fine-tuned with weak supervision to form the end-to-end model $h = g \circ f$. In many tasks, \ssl attains performance comparable to supervised learning\mcite{grill:2020:bootstrap}. 

Meanwhile, the popularity of \ssl also spurs intensive research on its security properties.
Existing work has explored the adversarial robustness of {\ssl}\mcite{jiang2020robust, fan2021does}. It is shown that, as a nice side effect, obviating the reliance on labeling may benefit the robustness to adversarial examples, label corruption, and common input corruptions\mcite{ssl-robustness, zhong2022self}. However, whether this robustness benefit also generalizes to other types of attacks remains an open question. This work explores this question in the context of backdoor attacks.

\subsection{Backdoor Attacks} As one major security threat, backdoor attacks inject malicious backdoors into the target model during training and activate such backdoors at inference. Typically, the backdoored model classifies trigger-embedded inputs to the adversary's designated class (effectiveness) but functions normally on clean inputs (evasiveness). Formally, under the supervised setting, the loss function of backdoor attacks is defined as:
\begin{equation}
\label{eq:trojan_sl}
\gL_\mathrm{bkd} = \sE_{(x, y) \in  \gD}\, \ell(h(x), y)  + \lambda \sE_{x_* \in  \gD_*}\, \ell(h(x_*), t)
\end{equation}
where $\ell$ represents the prediction loss, $\gD$ and $\gD_*$ respectively denote the clean and poisoning training data, $t$ is the target class designated by the adversary, and the hyper-parameter $\lambda$ balances the attack effectiveness and evasiveness.

Many backdoor attacks have been proposed for the supervised learning setting, which can be categorized along \mct{i} attack targets -- input-specific\mcite{poisonfrog}, class-specific\mcite{tact}, or any-input\mcite{badnet}, \mct{ii} attack vectors -- polluting training data\mcite{trojannn, xi2023security}, searching vulnerable architecture \mcite{pang2022dark} or releasing backdoored models\mcite{Ji:2018:ccsa}, and \mct{iii} optimization metrics -- effectiveness\mcite{imc}, transferability\mcite{latent-backdoor}, or  evasiveness\mcite{targeted-backdoor,poisonfrog,turner2019label,zhao2020clean}. 

Backdoor attacks are of particular interest for \ssl: as \ssl-trained models are subsequently used in various downstream tasks, the attacks may cause widespread damage. 
As most supervised backdoor attacks are inapplicable to \ssl due to their requirements for data labels, recent work has explored new ways of injecting backdoors into \ssl-trained models\mcite{jia:2021:badencoder, saha:2021:backdoor, liu:2022:poisonedencoder, carlini2021poisoning}: \mcite{carlini2021poisoning} focuses on the setting of multimodal contrastive learning;
BadEncoder\mcite{jia:2021:badencoder} injects backdoors into pre-trained encoders and releases backdoored models to victims in downstream tasks; SSLBackdoor\mcite{saha:2021:backdoor} generates poisoning data using a specific image patch as the trigger, while PoisonedEncoder\mcite{liu:2022:poisonedencoder} generates poisoning data by randomly combining target inputs with reference inputs. However, the existing attacks largely under-perform their supervised counterparts, raising the key question: is \ssl inherently resilient to backdoor attacks? 


\subsection{Backdoor Defenses}
To mitigate the threats of backdoor attacks, many defenses have been proposed, which can be categorized according to their strategies\mcite{trojanzoo}: \mct{i} input filtering, which purges poisoning examples from training data\mcite{tran:2018:nips,chen2018detecting}; \mct{ii} model inspection, which determines whether a given model is backdoored and, if so, recovers the target class and the potential trigger\mcite{kolouri2020universal, huang2019neuroninspect, abs, Wang:2019:sp}; and \mct{iii} input inspection, which detects trigger inputs at inference time\mcite{tact, gao2019strip, subedar2019deep}. However, designed for supervised backdoor attacks, the effectiveness of these defenses in the \ssl setting remains under-explored. 

\section{CTRL}
\label{sec:method}


\begin{figure}[t]
	\centering
	\includegraphics[width=0.85\columnwidth]{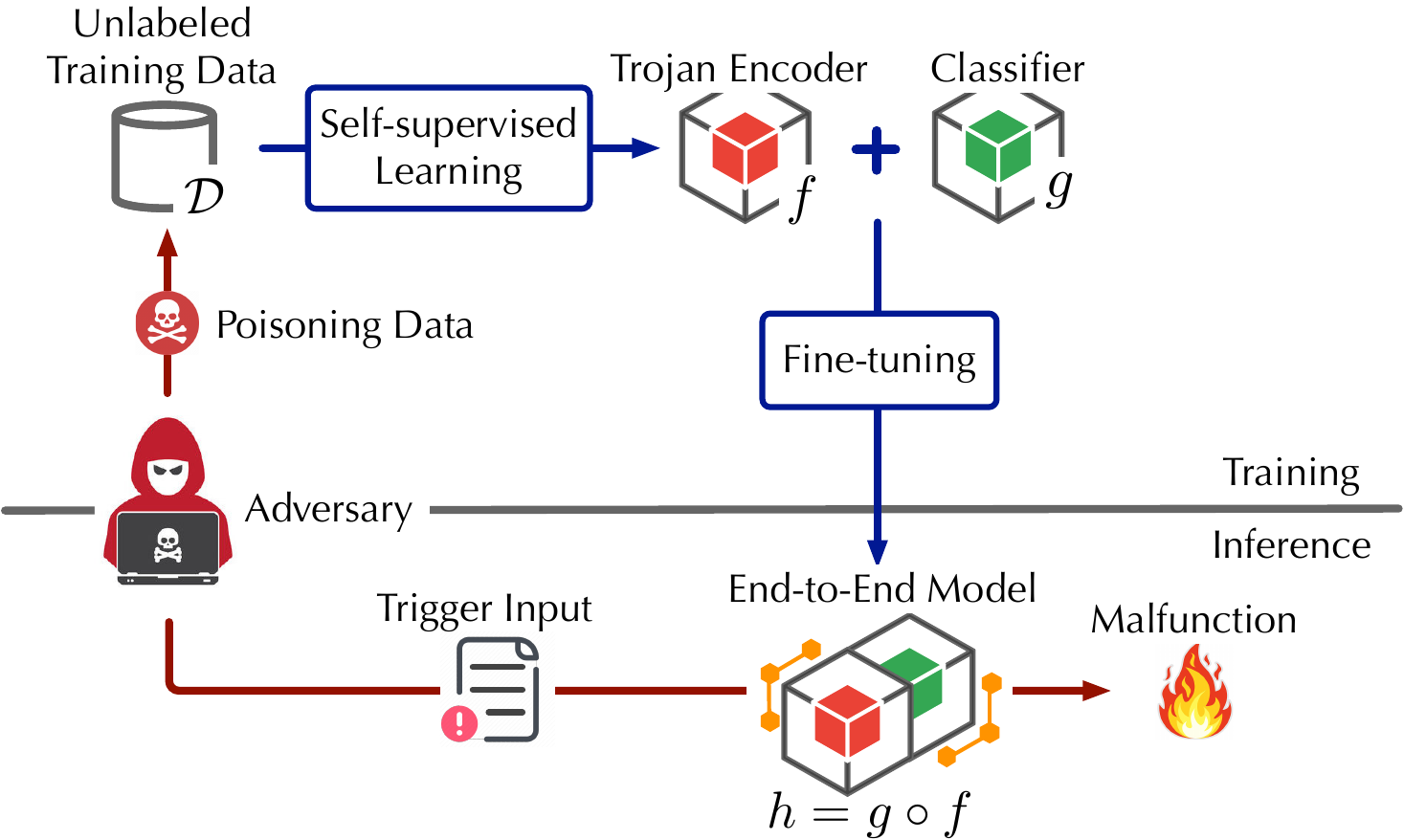}
	\caption{\small Illustration of self-supervised backdoor attacks. \label{fig:trojan}}
\end{figure}

In this section, we present \system, a simple yet effective self-supervised backdoor attack.

\subsection{Threat Model}

Following the existing work on trojan attacks\mcite{badnet,trojannn,imc}, we assume the threat model as illustrated in Figure\mref{fig:trojan}.

Attacker's objectives -- The adversary aims to inject malicious functions into the target model during training, such that at inference, any input embedded with a predefined trigger is classified into the adversary's target class while the model functions normally on clean inputs.


Attacker's capability -- The adversary attains the objectives by polluting a tiny fraction of the victim's training data. This assumption is practical for \ssl as it often uses massive unlabeled data collected from public data sources (\meg, Web), which opens the door for the adversary to pollute such sources and lure the victim into using poisoning data.

Attacker's knowledge -- We assume a black-box setting in which the adversary has no knowledge of \mct{i} the encoder and classifier models or \mct{ii} the training and fine-tuning regimes (\meg, classifier-only versus full-model tuning).

\subsection{Overview}
\label{sec:intuition}


Recall that \ssl \jiang{(with an emphasis on contrastive learning)} performs representation learning by optimizing the contrastive loss, which aligns the features of the same input under varying augmentations (``positive pair'') while separating the features of different inputs (``negative pair'') if applicable. The key idea of \system is three-fold: \mct{i} define the trigger as an augmentation-resistant perturbation, \mct{ii} generate poisoning data by adding the trigger to inputs from the target class, and \mct{iii} leverage the optimization of contrastive loss to entangle trigger inputs with target-class inputs in the feature space, which in turn leads to their similar classification in the downstream tasks.



We use a simplified model to explain the rationale behind \system. Let $x$ be a clean input from the target class. We assume the trigger embedding operator $\oplus$, which mixes $x$ with trigger $r$ to produce trigger input $x_* = x \oplus r$, can be disentangled in the feature space. That is, $f(x_*) = (1 -\alpha) f(x) + \alpha f(r)$, where $\alpha \in (0, 1)$ is the mixing weight.\footnote{In general, this property holds approximately for encoders that demonstrate linear mixability\mcite{mixup}.} With cosine similarity as the similarity metric, 
by aligning the positive pair $(x_*, x_*^\splus)$ of trigger input $x_*$, we have the following derivation:
\begin{equation}
\begin{split}
\label{eq:align}
\hspace{-15pt}
 f(x_*)^\sinv f(x_*^\splus)  = & 
       \underbrace{(1-\alpha)^2 f(x)^\sinv f(x^\splus)}_{\text{align clean inputs}}   + \underbrace{\alpha^2 f(r)^\sinv f(r^\splus)}_{\text{align triggers}} \\
     &  + \underbrace{\alpha(1-\alpha) (f(x)^\sinv f(r^\splus) + f(r)^\sinv  f(x^\splus))}_{\text{entangle trigger with target-class input}} 
       \end{split}
\end{equation}
where the first term aligns the positive pair of clean input $x$, the second term aligns trigger $r$ and its augmented variant, and the third term aligns $r$ with $x$. Observed that \mct{i} aligning the positive pair of trigger input $x_*$ naturally entangles trigger $r$ and target-class input $x$ in the feature space; \mct{ii} to maximize this entanglement effect, both $r$ and its variant $r^\splus$ need to be well represented in the feature space; in other words, the trigger pattern should be insensitive to varying augmentations (\meg, random cropping). More detailed analysis of this entanglement effect is deferred to \msec{sec:theory}.

\subsection{Implementation}
\label{sec:trigger}

Next, we elaborate on the implementation of \system based on the above insights.

\begin{figure}[!t]
	\centering
	\includegraphics[width=0.95\columnwidth]{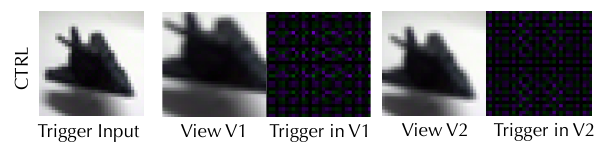}
	\caption{Illustration of the utilization of poisoning data (note: the trigger is magnified by 20 times to be evident). \label{figure:asy_sy}}
\end{figure}

{\bf Trigger definition --} To maximize the entanglement effect, we define trigger patterns as augmentation-resistant perturbations, \jiang{which means they are more likely to be retained after data augmentations in SSL.} Here, we use spectral triggers\mcite{spectral-trigger} as an example, which are specific perturbations in an input's frequency domain (\meg, increasing the magnitude of a particular frequency). Compared with other designs (\meg, image patches), spectral triggers are {\em augmentation-resistant} -- they are global (covering the entire input) and repetitive (periodic in the input's spatial domain), making them robust against augmentations, and {\em inspection-evasive} -- the perturbations on the input's high-frequency bands lead to visually invisible patterns. Intuitively, the perturbation frequency and magnitude are set to balance attack effectiveness and evasiveness, with lower frequency and larger magnitude leading to more effective (but less evasive) attacks. 
As shown in Figure\mref{figure:asy_sy}, the sample trigger is retained in various augmented views of the same input and invisible even with 20 times magnification. 

{\bf Poisoning data generation --} With trigger $r$, we generate poisoning data $\gD_*$ by applying $r$ to a set of candidate inputs. To this end, we assume the adversary has access to a small set of target-class inputs $\tilde{\gD}$. \deleted{Recall that the adversary has no control over which inputs to be used in training. To maximize the chance of poisoning data being selected,} In practice, the victim may only access and use a subset of poisoning data during training. To imitate this scenario, we randomly sample $k$ inputs from $\tilde{\gD}$ as candidates to craft  $\gD_*$.

\begin{figure}[!ht]
	\centering
	\includegraphics[width=0.8\columnwidth]{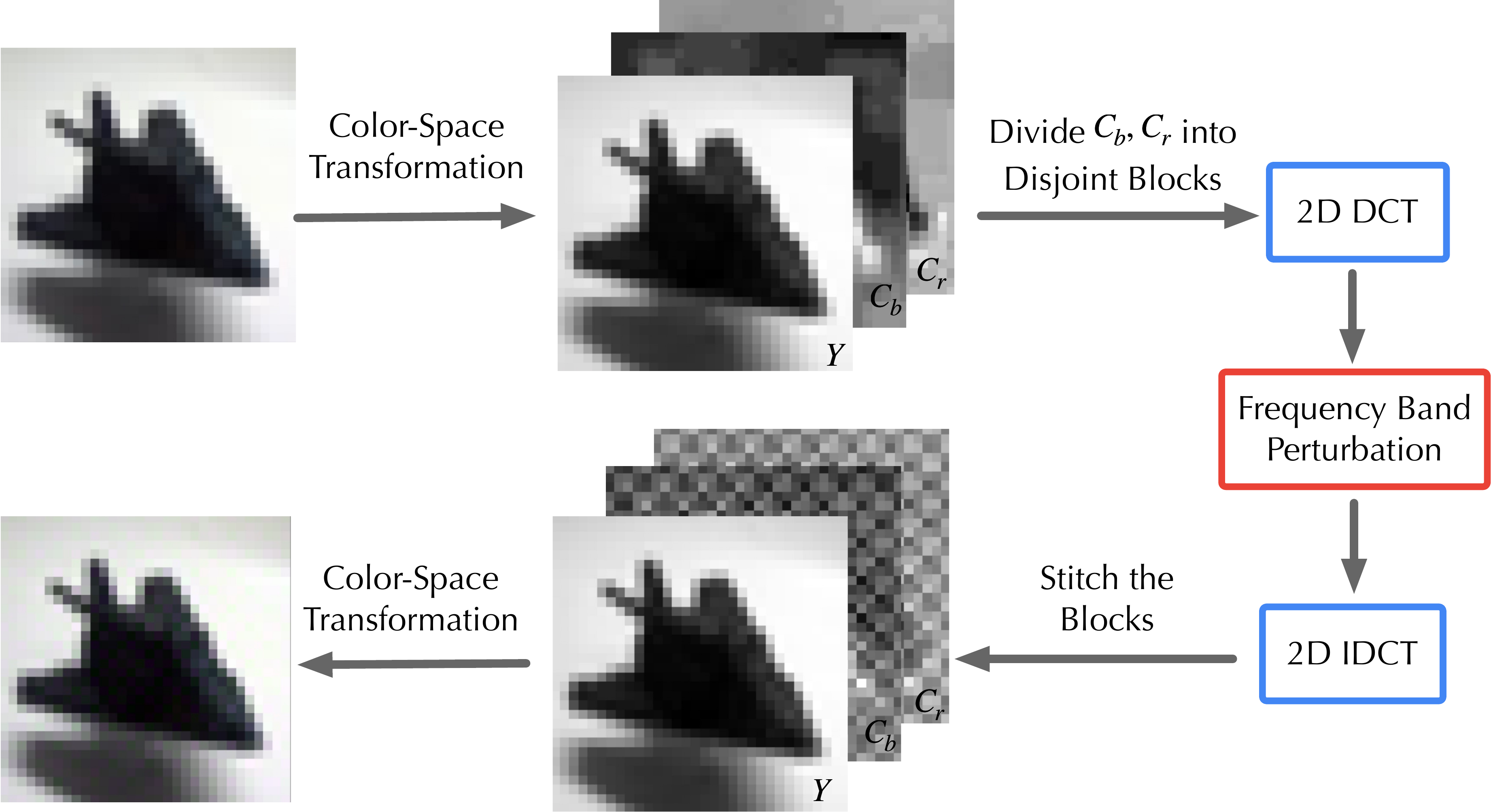}
	\caption{\small Illustration of the process of trigger generation. \label{fig:pattern}}
\end{figure}

{\bf Trigger embedding and activation --} To embed trigger $r$ into given input $x$, we first convert $x$ to the YC$_\text{b}$C$_\text{r}$ color space, 
which separates $x$'s luminance component (Y) from its chrominance component (C$_\text{b}$ and C$_\text{r}$). As human perception is insensitive to chrominance change\mcite{hemalatha2013comparison}, we apply perturbation to the C$_b$ and C$_r$ channels only. Specifically, we use discrete cosine transform (DCT)\mcite{rahman2013dwt} to transform $x$ to the frequency domain 
and apply the perturbation defined in $r$. 
We then use inverse DCT to transform $x$ back to the spatial domain 
and convert it to the RGB color space to form the trigger input. This process is illustrated in Figure\mref{fig:pattern}.

Note that the spectral trigger definition makes it possible to decouple the setting of triggers for crafting poisoning data (\mie, small magnitude to optimize the attack evasiveness) and activating backdoors at inference (\mie, large magnitude to optimize the attack effectiveness).

\section{Evaluation}
\label{sec:eval}




\vspace{-1em}
\subsection{Experimental Setting}

We begin by introducing the main setting of our evaluation. More details are deferred to Appendix\,\msec{app:setting}.

{\bf Datasets --} Our evaluation primarily uses three benchmark datasets: CIFAR-10\mcite{cifar} consists of 32$\times$32 color images in 10 classes; CIFAR-100\mcite{cifar100} is similar to CIFAR-10 but includes 100 classes; ImageNet-100 is a subset sampled (re-scaled to 64$\times$64) from the ImageNet-1K dataset\mcite{Deng:2009:cvpr} and contains 100 randomly selected classes. 
Under the transfer setting, we also use GTSRB, which contains 32$\times$32 traffic-sign images in 43 classes, as an additional dataset.


{\bf Metrics --} 
We mainly use two metrics: attack success rate (\asr) measures the accuracy of the model in classifying trigger inputs as the adversary's designated class, while clean data accuracy (\acc) measures the accuracy of the model in classifying clean inputs. In the transfer setting, we evaluate untargeted attacks by measuring the model's accuracy drop on trigger inputs.
{\bf SSL methods --} We mainly use three representative contrastive learning methods,  SimCLR\mcite{chen:2020:simple}, BYOL\mcite{grill:2020:bootstrap}, and SimSiam\mcite{chen:2021:exploring}.  Their accuracy on the benchmark datasets is summarized in Table\mref{clean_acc} in Appendix\,\msec{app:details}. 

{\bf Models --} By default, we use an encoder with ResNet-18\mcite{resnet} as its backbone and a two-layer MLP projector to map the representations to a 128-dimensional latent space; further, we use a two-layer MLP with the hidden-layer size of 128 as the downstream classifier. We also explore alternative architectures in \msec{sec:sensitive}. Following prior work\mcite{chen:2020:simple, chen:2021:exploring}, we use \{RandomResizeCrop, RandomHorizontalFlip, ColorJitter, RandomGrayscale\} as the set of augmentations.

\begin{figure}[!ht]
	\centering
	\includegraphics[width=0.9\columnwidth]{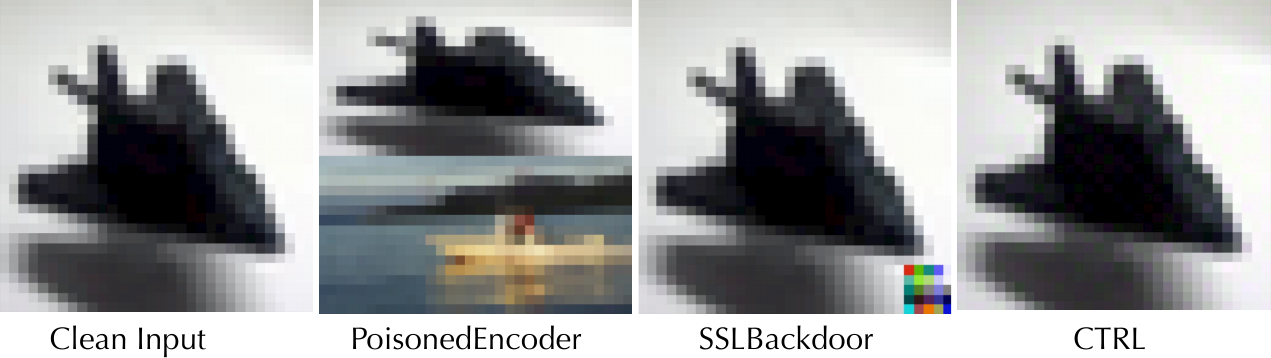}
	\caption{Comparison of the poisoning data of different attacks. \label{figure:perturb}}
\end{figure}

{\bf Attacks --} Given the limited prior work on self-supervised backdoor attacks, we compare \system with two baselines given their similar threat models: 
{\sslbkd}\mcite{saha:2021:backdoor} defines the trigger as a randomly positioned image patch (\meg, 5$\times$5);
{\bdenc}\mcite{liu:2022:poisonedencoder} targets specific inputs and combines target inputs with reference inputs to generate poisoning data; \system defines the trigger as increasing the magnitude of selected frequency bands of given inputs. By default, we set the perturbation frequency as 15 and 31 and the perturbation magnitude as 50 for generating poisoning data and 100 for activating backdoors at inference time. Figure\mref{figure:perturb} compares the poisoning samples generated by different attacks. Observe that compared with other attacks, the poisoning samples of \system are highly indistinguishable from clean data, leading to its evasiveness with respect to input inspection.

\subsection{Attack Effectiveness}



\begin{table}[ht]{\small
\renewcommand{\arraystretch}{1.0}
\setlength{\tabcolsep}{1.2pt}
\centering

\begin{tabular}{c|c|cccccc}
\multirow{3}{*}{Attack}      & \multirow{3}{*}{Dataset}  & \multicolumn{6}{c}{SSL Method}\\
\cline{3-8}
& & \multicolumn{2}{c}{SimCLR}   & \multicolumn{2}{c}{BYOL}   & \multicolumn{2}{c}{SimSiam}  \\ \cline{3-8}
                             &                          & ACC      & ASR                & ACC      & ASR              & ACC      & ASR                                       \\ \hline \hline
\multirow{3}{*}{\bdenc}         & CIFAR-10                  &   80.5\%      &    11.1\%                &  81.7\%       &     10.7\%             &     81.9\%    &        10.7\%                                      \\
                             & CIFAR-100                 &   47.9\%      &    1.3\%                &   50.9\%      &    1.2\%              &   52.3\%      &   1.2\%                                           \\
                             & ImageNet-100                 &     41.9\%    &       1.0\%             &    44.8\%     &   1.4\%               &      41.5\%    &         1.3\%                                  \\ \hline
\multirow{3}{*}{\sslbkd}  & CIFAR10                  &    79.4\%     &      33.2\%              &     80.3\%  &   46.2\%               &    80.6\%    &          53.1\%                                   \\
                             & CIFAR-100                 &      46.3\%    &        4.2\%            &      49.4\%   &       6.3\%           &    50.7\%  &        4.9\%                                       \\
                             & ImageNet-100                    &    40.7\%     &      10.2\%              &     43.3\%    &  7.6\%                &     38.9\%    &     5.5\%                                    \\ \hline
\multirow{3}{*}{\minitab[c]{\sslbkd \\(fixed)}}    &                         CIFAR-10                  &    80.0\%     &      10.5\%              &     82.3\%  &   11.2\%               &    81.9\%    &          10.7\%                                   \\
& CIFAR-100                 &      48.3\%    &        1.2\%            &      50.4\%   &       1.2\%           &    52.2\%  &        1.2\%                                       \\
& ImageNet                    &    42.0\%     &      1.1\%              &     45.4\%    &  1.2\%                &     41.2\%    &     1.3\%                                    \\ \hline
\hline
\multirow{3}{*}{\system}         & CIFAR-10                  &  80.5\% & \cellcolor{Red} 85.3\%    & 82.2\% & \cellcolor{Red} 61.9\%  &  82.0\% & \cellcolor{Red} 74.9\%                           \\
                             & CIFAR-100                 & 47.6\% & \cellcolor{Red} 68.8\%  & 50.8\%  & \cellcolor{Red} 42.3\% &  52.6\% & \cellcolor{Red} 83.9\%                           \\
                             & ImageNet-100&   42.2\%   &  \cellcolor{Red} 20.4\%  &  45.9\% &\cellcolor{Red}  37.9\% & 40.2\% & \cellcolor{Red} 39.2\%

\end{tabular}
\caption{Effectiveness of \system and baseline attacks.}
\label{attack_effect}}
\end{table}

\begin{figure}[!t]
	\centering
	\includegraphics[width=0.9\columnwidth]{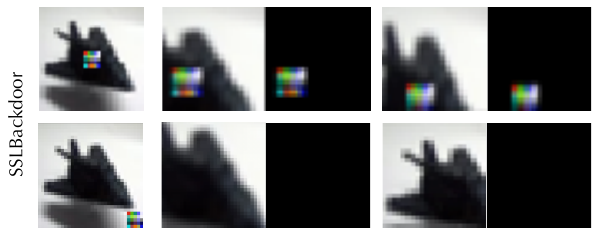}
	\caption{Utilization of poisoning data generated by \sslbkd. where the upper row shows the case of the center-positioned trigger; the lower row shows the case of the corner-positioned trigger. \label{figure:sslbackdoor}}
\end{figure}


  
\label{sec:target_attack}
  
{\bf Targeted attacks --} We evaluate the effectiveness of different attacks against representative \ssl methods on benchmark datasets. For a fair comparison, we fix the poisoning ratio of all the attacks as 1\%. The results are summarized in Table\mref{attack_effect}. We have the following observations. \mct{i} Across all the settings, \system attains the highest attack effectiveness. For instance, it achieves 83.9\% \asr (higher than the model's \acc) on CIFAR-100 when the backdoored model is trained using SimSiam. 
\mct{ii} In comparison, \sslbkd is much less effective, which may be attributed to its trigger design: as shown in Figure\mref{figure:sslbackdoor}, defined as a randomly positioned image patch, the trigger pattern can be easily distorted by augmentations, resulting in poor utilization of poisoning data.
To validate this hypothesis, we fix the trigger at the lower-right corner of an image (\sslbkd fixed); as shown Table\mref{attack_effect}, the \asr of \sslbkd (fixed) is close to random guess.
\mct{iii}  Meanwhile, the effectiveness of \bdenc is also limited. Recall that it only targets specific inputs and generates poisoning data by combining target inputs with reference inputs (\mcf Figure\mref{figure:perturb}), thereby being unable to generalize to all trigger-embedded inputs.

\textbf{
}\begin{figure}[!ht]
\centering
  \includegraphics[width=0.85\columnwidth]{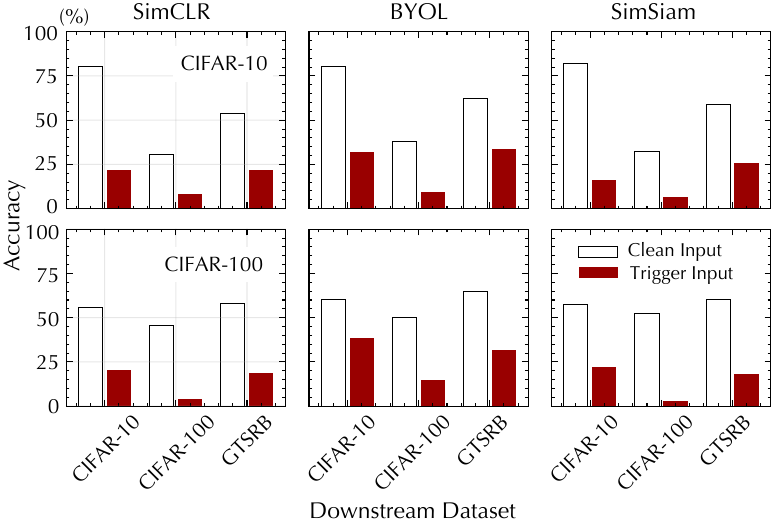}%
\caption{Model accuracy of classifying clean and trigger input under the transfer setting.} 
\label{figure:acc_transfer}
\end{figure}

{\bf Untargeted attacks --}
In the transfer scenario, the victim trains an encoder on a pre-training dataset using \ssl and then fine-tunes the downstream classifier using another dataset. \deleted{Next, we measure the attack effectiveness under this setting.  }
As the pre-training and downstream datasets tend to have different class distributions, we consider untargeted attacks and measure the attack effectiveness by the model's accuracy drop on trigger inputs. 
Figure\mref{figure:acc_transfer} shows the backdoored model's accuracy of classifying clean and trigger inputs when the pre-training dataset is CIFAR-10 or CIFAR-100. Specifically, even if the pre-training and downstream datasets are different, \system greatly decreases the model's accuracy in classifying trigger inputs. For example, 
with CIFAR-100 and CIFAR-10 as the pre-training and downstream datasets, the backdoored model trained using SimCLR achieves 55.7\% and 20.5\% accuracy on clean and trigger inputs, respectively. \deleted{This large gap indicates a highly effective untargeted attack.} Further, we find that even if the downstream dataset does not contain the adversary's target class, the trigger inputs tend to be misclassified to certain classes (\meg,  ``bus'', ``pickup truck'', and ``train'' in the downstream dataset) that are semantically similar to the target class (\meg,  ``truck'' in the pre-training dataset).



\subsection{Sensitivity Analysis}
\label{sec:sensitive}

We next explore the sensitivity of \system to external factors including encoder models and fine-tuning methods. The results of other factors are deferred to Appendix \msec{app:details}.

\begin{table}[!ht]{\small
\centering
\renewcommand{\arraystretch}{1.1}
\begin{tabular}{rcc}
Encoder Model         & ACC & ASR \\ \hline
ResNet-18     &  80.5\%   &  85.3\%   \\
MobileNet-V2  &  76.4\%  &   79.8\%   \\
SqueezeNet    &  74.7\%  &    54.8\% \\
ShuffleNet-V2 &  76.2\%  &   38.3\%  \\ \hline
\end{tabular}
\caption{Evaluation on different model architectures.}
\label{table:arch}}
\end{table}

{\bf Encoder models --} The previous experiments are conducted on an encoder with ResNet-18 as its backbone. We now evaluate the impact of the encoder model on the performance of \system on CIFAR-10. We evaluate the \acc and \asr of \system on encoders of various architectures including ShuffleNet-V2\mcite{shufflenet}, MobileNet-V2\mcite{mobilenetv2}, and SqueezeNet\mcite{squeezenet}, with the other settings fixed the same as Table\mref{attack_effect}. 
As shown in Table\mref{table:arch}, \system attains high \asr across all the other architectures (\meg, 79.8\% \asr on MobileNet-V2), indicating its insensitivity to the encoder model.




 \deleted{Further, observe that both \acc and \asr of \system are slightly lower on MobileNet, SqueezeNet, and ShuffleNet, compared with ResNet, implying a possible accuracy-robustness trade-off\mcite{robust-accuracy}. We consider investigating this trade-off as our ongoing research.}




{\bf Fine-tuning methods --} In fine-tuning the downstream classifier, the victim may opt to use different strategies (\meg, classifier-only versus full-model tuning). Recall that the adversary has no knowledge about fine-tuning. We evaluate the impact of the fine-tuning method on the attack performance. Table\mref{table:tune} summarizes the \acc and \asr of \system on CIFAR-10 under SimCLR with varying fine-tuning strategy and trigger magnitude. We have the following observations. 


\begin{table}[!ht]{\small
\centering
\renewcommand{\arraystretch}{1.1}
\setlength{\tabcolsep}{3pt}
\begin{tabular}{cccc}
Trigger Magnitude            & Fine-tuning Method          & ACC              & ASR                \\ \hline
\multirow{2}{*}{50}  & classifier-only & 80.6\%  & 67.3\%    \\
                     & full-model      &  84.4\%   &  65.1\%                   \\
                     \hline
\multirow{2}{*}{100} & classifier-only & 81.1\%  & 86.3\%    \\
                     & full-model      & 84.5\%   & 71.7\%   \\ 
\end{tabular}
\caption{Performance of \system w.r.t. fine-tuning strategy and trigger magnitude on CIFAR-10 under SimCLR. }
\label{table:tune}}
\end{table}

First, compared with classifier-only tuning, fine-tuning the full model improves the model accuracy. For instance, with the trigger magnitude set as 50, full-model tuning improves the \acc from 80.6\% to 84.4\%. Second, the fine-tuning strategy has a modest impact on the \asr of \system. For instance, with the trigger magnitude set as 50, the \asr under classifier-only and full-model tuning differs by only 2.2\%. Finally, increasing the trigger magnitude generally improves \asr under varying fine-tuning strategies. For instance, it grows by 6.6\% under full-model tuning if the trigger magnitude increases from 50 to 100.

\subsection{Ablation Study}

Below we conduct an ablation study to understand the contribution of each component of \system to its effectiveness.

{\bf Candidate selection --} 
Besides randomly selecting candidate inputs to craft poisoning data in \msec{sec:trigger}, we consider alternative scenarios: {\em center} -- we train a clean encoder $f$ on the reference data $\tilde{\gD}$, compute the representation of each input in $\tilde{\gD}$, and then select $k$ candidates closest to the center in the feature space (measured by $L_2$ distance); and {\em core-set} -- we cluster the inputs in $\tilde{\gD}$ into $k$ clusters in the feature space (\meg, using $k$-means clustering) and select the inputs closest to the cluster centers as the candidates.

\begin{table}[!ht]{\small 
\renewcommand{\arraystretch}{1.1}
\setlength{\tabcolsep}{1pt}
\centering
\begin{tabular}{c|c|cccccc}
\multirow{2}{*}{Dataset}  & \multirow{2}{*}{Selector} & \multicolumn{2}{c}{SimCLR} & \multicolumn{2}{c}{BYOL} & \multicolumn{2}{c}{SimSiam} \\ \cline{3-8} 
                          &                            & ACC           & ASR         & ACC          & ASR        & ACC           & ASR          \\ \hline
\multirow{3}{*}{CIFAR-10}  & Random                     & 80.5\%      & \cellcolor{Red} 85.3\%      &  82.2\%     & \cellcolor{Red} 61.9\%     &  82.0\%      & \cellcolor{Red} 74.9\%      \\
                          & Center                     & 81.2\%      & 57.1\%     & 80.0\%     & 47.4\%    & 80.8\%      & 67.9\%       \\ 
                          & Core-set                     &  80.3\%     &  31.2\%   &  81.7\%    &  52.6\%  &   81.7\%   & 40.4\%     \\ \hline
\multirow{3}{*}{CIFAR-100} & Random                     & 47.6\%      & 68.8\%     & 50.8\%     & 42.3\%    & 52.6\%      & \cellcolor{Red} 83.9\%      \\
                          & Center                     & 48.8\%      & \cellcolor{Red}78.1\%     & 51.2\%     & 54.7\%    & 52.8\%      & 78.6\%      \\ 
                          & Core-set                     &   48.5\%    &  54.6\%   &  50.7\%    & \cellcolor{Red}64.7\%    &  53.1\%    &  53.9\%      \\ \hline
\end{tabular}
\caption{Performance of \system with varying candidate selectors.}
\label{example_selection}}
\end{table}

Table\mref{example_selection} compares their impact on the attack performance. First, the random selector outperforms others on CIFAR-10. This may be explained by that over the relatively simple class distribution (\meg, 10 classes), the random scheme is able to select a set of representative candidates of the underlying distribution. Second, no single selector dominates on CIFAR-100. This may be explained by that no single selector is able to fit the complex class distribution (\mie, 100 classes) across all the \ssl methods. Thus, under the setting where the adversary can poison a limited amount of training data, the random selector is a practical choice.

\begin{figure}[!ht]
\centering
\includegraphics[width=0.85\columnwidth]{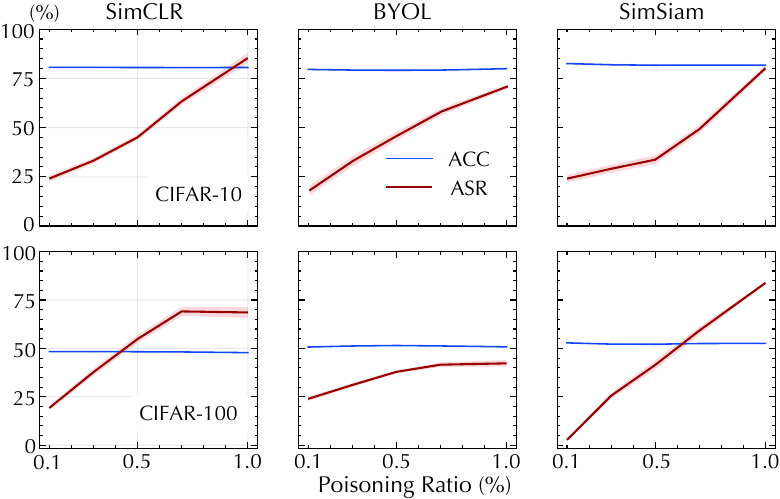}%

\caption{Performance of \system with respect to poisoning ratio.}
\label{figure:poison_ratio}
\end{figure}

{\bf Poisoning ratio --} 
In Figure\mref{figure:poison_ratio}, we show that increasing the poisoning ratio from 0.1\% to 1\% has little effect on the model's performance on clean inputs, but significantly increases the attack effectiveness of \system. For example, on CIFAR-10 with SimCLR, increasing the poisoning ratio from 0.1\% to 1\% leads to a 61.2\% increase in \asr. Further, even with a 0.5\% poisoning ratio (100 out of 50,000 training samples), \system is still able to inject effective backdoors into the models (close to 50\% \asr on CIFAR-10 with SimCLR), indicating its practicality in the real-world scenarios. 

\begin{table}[h]{\small
\renewcommand{\arraystretch}{1.1}
\setlength{\tabcolsep}{3pt}
\centering
\begin{tabular}{c|ccccl}
Poisoning Ratio & 0.1\% & 0.3\% & 0.5\% & 0.7\% & 1\%  \\ \hline
ASR          & 12\%  & 34\%  & 59\%  & 64\%  & 69\% \\ \end{tabular}
\caption{ASR with respect to the poisoning ratio on CIFAR-100 (SimCLR). \label{ratio}}
}
\end{table}

\jiang{Additionally, we explore the attack effectiveness of \system with varying poisoning ratios on CIFAR-100.  As shown in Table\,\ref{ratio}, CTRL attains high ASR under a low poisoning ratio on CIFAR-100 (which can be further enhanced by adjusting the trigger strength at inference). For example, the poisoning ratio of 0.1\% can achieve an ASR of 12\%.}

\begin{figure}[t]
\centering
 \includegraphics[width=0.85\columnwidth]{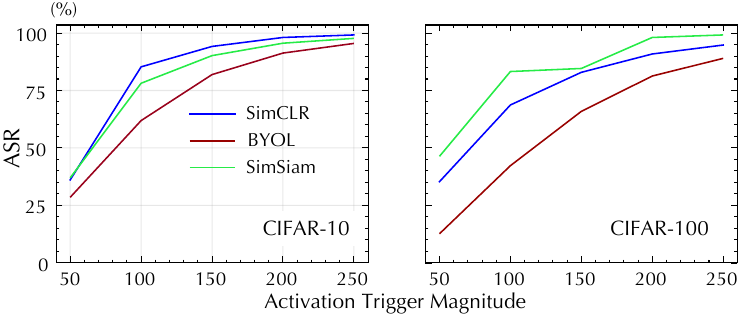}%
\caption{\asr of \system  with respect to activation trigger magnitude on CIFAR-10 with SimCLR. }
\label{fig:act_magnitude}
\end{figure}

{\bf Backdoor activation --} 
Recall that \system allows the adversary to set triggers of different magnitude for poisoning trigger and activation trigger.
We evaluate the influence of the activation trigger magnitude (with the poisoning trigger magnitude fixed as 50) on the attack performance, with results summarized in Figure\mref{fig:act_magnitude}. As expected, increasing the activation trigger magnitude improves the attack effectiveness. For instance, on CIFAR-10 with SimCLR, as the activation trigger magnitude varies from 50 to 250, the \asr of \system increases from 36\% to 99\%. Note that as the activation trigger is only applied at inference, increasing the activation trigger magnitude does not affect the \acc.

\section{Discussion}
\label{sec:explore}

Thus far, we show empirically that \ssl is highly vulnerable to backdoor attacks. Next, through the lens of \system, we study the potential root of this vulnerability and its implications for defenses. 

\subsection{Characterizing Self-supervised Backdoor Attacks}
\label{sec:theory}

In \msec{sec:intuition}, we give an intuitive explanation about how \system leverages the optimization of contrastive loss to entangle trigger-embedded and target-class inputs in the feature space. We now quantitatively characterize this entanglement effect. Specifically, under the alignment and uniformity assumptions commonly observed in \ssl-trained encoders\mcite{alignment}, we have the following theorem (proof in Appendix\,\msec{sec:proofs}):

\vspace{-0.5em}

\begin{theorem}
\label{the:attack_main}
Let $\tilde{x}$ be a clean input randomly sampled from a non-target class and $x$ be a clean input randomly sampled from the target class $t$. The entanglement between the trigger-embedded input $\tilde{x}_* = \tilde{x} \oplus r$ and $x$ in the feature space is lower bounded by: $\sE[f(\tilde{x}_*)^\intercal f(x)] \geq \alpha - \frac{\epsilon}{2(1- \alpha)}$, where $\alpha$ is the mixing weight in \meq{eq:align}, \jiang{and $\epsilon \in [0, 1)$ is a small non-negative number.}
\end{theorem}

Theorem\mref{the:attack_main} shows that the entanglement is not a monotonic function of $\alpha$: with overly small $\alpha$, the influence of the trigger pattern on the entanglement is insignificant; with overly large $\alpha$, the trigger pattern dominates the features of trigger-embedded inputs, which also negatively impacts the entanglement effect.

To validate our analysis, we empirically measure the entanglement effect between trigger-embedded and target-class inputs by varying the trigger magnitude in \system (\mcf \msec{sec:trigger}). Specifically, we define entanglement ratio (ER), a metric to measure the entanglement effect, which extends the confusion ratio metric used in\mcite{wang:2022:chaos} to our setting. We sample $n = 800$ clean inputs from each class of CIFAR-10 to form the dataset $\gD$; we apply a set of $m = 10$ augmentation operators $\gA$ (sampled from the same distribution used by SSL) to each input $x \in \gD$, which generates an augmented set $\gD^\splus = \{ a(x) \}_{ x\in \gD, a \in \gA}$. Further, we randomly sample $1,000$ clean inputs disjoint with $\gD$ across all the classes and generate their trigger-embedded variants $\gD_*$. For each $x_* \in \gD_*$, we find its $K = 100$ nearest neighbors $\gN_{K}(f(x_*))$ among $\gD^\splus$ in the feature space and then measure the proportion of neighbors from the target class $t$:
\begin{equation}
\label{eq:metric}
  \mathrm{ER}(f) = \frac{1}{K}  \sE_{x_* \in  \gD_*}\mathbbm{1}_{  f(x^\splus)  \in  \gN_{K}(f(x_*)), c(x^\splus) = t}
\end{equation}
where $t$ is the target class, $\mathbbm{1}_p$ is an indicator function that returns 1 if $p$ is true and 0 otherwise, $c(x^\splus)$ returns $x^\splus$'s label. Note that we use the label information here only for understanding the entanglement effect. 


\begin{figure}[!t]
	\centering
	\includegraphics[width=0.85\columnwidth]{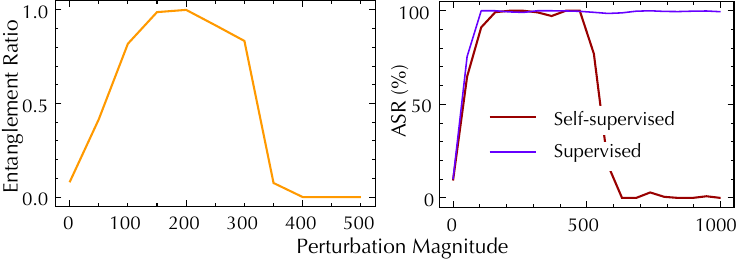}
	\caption{Entanglement effect and \asr with respect to trigger magnitude. \label{fig:overlap}}
\end{figure}




Intuitively, a larger \er indicates a stronger entanglement effect between trigger-embedded and target-class inputs. We measure \er under varying trigger magnitude, with results shown in Figure\mref{fig:overlap}. With the increase of trigger magnitude (a proxy of $\alpha$), the entanglement effect first grows from 0 to 100\% and then drops gradually to 0, which is consistent with our theoretical analysis. 

Now, we show that this entanglement effect may account for the effectiveness of \system. Figure\mref{fig:overlap} measures the \asr of \system under varying trigger magnitude (the same magnitude for the poisoning and activation triggers).
Observe that \asr demonstrates a trend highly similar to \er with respect to trigger magnitude: it first increases to 100\% and then drops to 0. Also notice that the trend of \asr lags behind \er. This may be explained as follows: the classifier divides the feature space into different classes; only when trigger-embedded and target-class inputs are separated sufficiently apart, the \asr starts to drop. In comparison, the \asr of supervised trojan attack increases to around 100\% and maintains at that level, indicating its irrelevance to the entanglement effect.
This observation implies that it is critical to optimally tune the entanglement effect to maximize the attack effectiveness.

\subsection{Adversarial Robustness versus Backdoor Vulnerability}

Prior work shows that \ssl may benefit the robustness to adversarial perturbation, label corruption, and data distribution shift\mcite{ssl-robustness,  zhong2022self, wu2023bottrinet, li2019det, qiu2022hijack}. However, our empirical evaluation and theoretical analysis suggest that this robustness benefit may not generalize to backdoor attacks. We speculate that the representation-invariant property of \ssl, which benefits such robustness, may also be the very reason making \ssl vulnerable to backdoor attacks. 

Intuitively, representation invariance indicates that different augmented views of the same input should share similar representations. Essentially, data augmentation and contrastive loss, two key ingredients of \ssl, are designed to ensure this property\mcite{chen:2020:simple, grill:2020:bootstrap, chen:2021:exploring}. Meanwhile, robustness indicates that some variants of the same input should share the same label (\mie, label invariance). Thus, these two properties are aligned in principle; enforcing the invariance of intermediate representations tends to improve the variance of classification labels.

On the other hand, due to the entanglement between the augmented views of trigger-embedded and target-class inputs, enforcing the representation invariance causes the trigger-embedded and target-class inputs to generate similar representations and essentially entangles them in the feature space, leading to the risk of backdoor attacks. Therefore, the robustness of \ssl to adversarial attacks may be at odds with its robustness to backdoor attacks.


\subsection{Defense Challenges}

The entanglement between the representations of trigger and clean inputs also causes challenges for defenses that rely on the separability of trigger inputs. Here, we explore such challenges using several state-of-the-art defenses.

\begin{figure}[!ht]
	\centering
	\includegraphics[width=0.75\columnwidth]{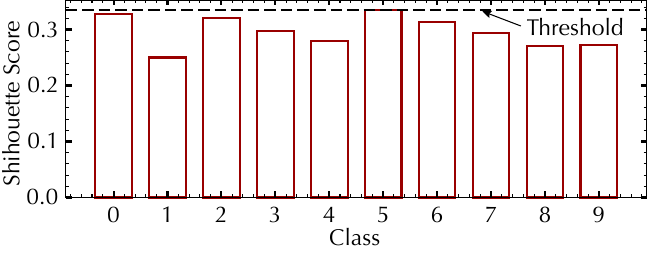}
	\caption{Evaluation results of activation clustering against \system. \label{fig:ac}}
\end{figure}

{\bf Activation clustering (AC) --} Based on the premise that in the target class, poisoning samples form their own cluster that is small or far from the class center, AC detects the target class using the silhouette score of each class\mcite{chen2018detecting}. 
Due to its reliance on labeling, AC is inapplicable to \ssl directly. Here, we assume labels are available and explore its effectiveness against \system on CIFAR-10 with SimCLR. 
From Figure\mref{fig:ac}, observe that AC fails to identify the target class (class 0), which has a lower score compared to other classes (\meg, class 5), not to mention detecting poisoning inputs. 


{\bf Statistical contamination analyzer (SCAn) --} It detects trigger inputs based on the statistical anomaly of their representations.
Following\mcite{tact}, we randomly sample 1,000 inputs from the testing set to build the decomposition model; we use it to analyze a poisoning set with 5,000 trigger inputs and 5,000 clean inputs. We use FPR and TPR to evaluate SCAn. To compare the performance of SCAn against supervised and self-supervised backdoor attacks, we also evaluate SCAn on two supervised backdoor attacks: one with the same spectral trigger as \system and the other with a random 5$\times$5 image patch as the trigger. Table\mref{table:scan} summarizes the results. We have the following key observations. First, it is more challenging for SCAn to detect spectral triggers than patch triggers. For instance, with FPR fixed as 0.5\%, the TPR of SCAn differs by over 34\% on the spectral and patch triggers. The difference may be explained by that compared with patch triggers, spectral triggers are more evasive by design (\mcf \msec{sec:method}), which can hardly be characterized by a mixture model. Even if the target class is correctly identified, many trigger inputs may still fall into the cluster of clean inputs. 

\begin{table}[!ht]{\small 
\centering
\renewcommand{\arraystretch}{1.0}
\begin{tabular}{c|ccc}
\multirow{2}{*}{FPR} & \multicolumn{3}{c}{TPR}    \\
\cline{2-4}
                     & \system & Supervised (spectral) & Supervised (patch) \\ \hline
0.5\%                    & 28.0\%    & 63.0\%        & 97.0\%        \\
1.0\%                & 28.0\%    & 66.5\%          & 97.0\%        \\
2.0\%                  & 28.0\%    & 68.0\%        & 97.0\%        \\ 
\end{tabular}
\caption{Evaluation results of SCAn against \system. \label{table:scan}}}
\end{table}

{\bf Robust training --} Adding limited Gaussian noise to the training data tends to improve the model robustness while maintaining the performance on the original task\mcite{zheng2016improving,li2021towards}.  Following\mcite{zheng2016improving}, we add noise to the training data as a possible defense. Our results show that \system maintains high ASR with noise levels up to 16/255. When the noise level is further increased to 25/255, the ASR drops to 16\%, leading to 2.1\% accuracy (ACC) drop. We attribute this to the use of a small magnitude trigger to maintain the attack's stealthiness, which can be disrupted by strong Gaussian noise. Nonetheless, determining the optimal magnitude of defensive noise poses a challenge as the defender is not privy to the specifics of the trigger, making it challenging to strike a balance between ACC and defense effectiveness.

\jiang{{\bf Other defenses --} We examine several additional defenses. MNTD may be infeasible for SSL due to its requirement of training a large number of shadow models (e.g., 4,096 clean/trojan) \mcite{xu2021detecting}. NeuralCleanse \cite{wang2019neural}, a trigger inversion defense, fails in all trials with an anomaly index averaging 0.72 ± 0.38 (below the threshold of 2). We leave the exploration on more other defenses as future work.}

\subsection{Limitations}


Next, we discuss the limitations of this work. First, existing work\mcite{saha:2021:backdoor} has already studied self-supervised backdoor attacks. However, this work significantly improves the SOTA attack success rate of self-supervised backdoor attacks, suggesting that SSL is comparably vulnerable to backdoor attacks as supervised learning.
Moreover, we identify the underlying differences between the mechanisms of \ssl and supervised backdoor attacks, enabling us to extend our approach to other trigger definitions. Second, we define the trigger based on heuristics, which is not necessarily optimal. We mainly use it as an example to study the unique vulnerability of \ssl. How to rigorously optimize the trigger design of \system represents an intriguing question. Finally, we mainly focus on image classification tasks, while \ssl has been applied in many other domains, such as natural language processing and graph learning. We consider extending \system to such domains as ongoing work.

\section{Conclusion}
\label{sec:conclusion}
This work conducts a systematic study on the vulnerability of self-supervised learning (\ssl) to backdoor attacks. By developing and evaluating \system, a simple yet highly effective self-supervised backdoor attack, which dramatically bridges the gap in the attack effectiveness of backdoor attacks between \ssl and supervised counterparts.
Further, both empirically and analytically, we reveal that the representation invariance property of \ssl, which benefits adversarial robustness, may also account for this vulnerability. Finally, we discuss the unique challenges to defending against self-supervised backdoor attacks. We hope our findings will shed light on developing more robust \ssl methods.

\vspace{2em}
{\bf \noindent Acknowledgment:}  We thank the anonymous reviewers for their valuable feedback. This work is supported by the National Science Foundation under Grant No. 1951729, 1953893, 2119331, and 2212323. 

\newpage
\bibliographystyle{plain}
\bibliography{bibs/aml,bibs/optimization,bibs/general,bibs/ting,bibs/graph, bibs/ssl, bibs/interpretation}

\clearpage
\appendix

\section{Characterizing Supervised Backdoor Attacks}
\label{sec:sl_backdoor}

In supervised learning, the backdoor attack associates the trigger $r$ with the target label $t$ via (implicitly) minimizing the objective defined in \meq{eq:trojan_sl}.
The success of this attack is often attributed to the model's excess capacity\mcite{excess-capacity}, which ``memorizes'' both the function that classifies clean inputs and that misclassifies trigger inputs. 
Note that \meq{eq:trojan_sl} does not specify any constraints on the representations of trigger inputs. Thus, 
while associated with the same class, the trigger-embedded and target-class inputs are not necessarily proximate in the feature space.  

\begin{figure}[ht]
\centering
  \includegraphics[width=0.7\columnwidth]{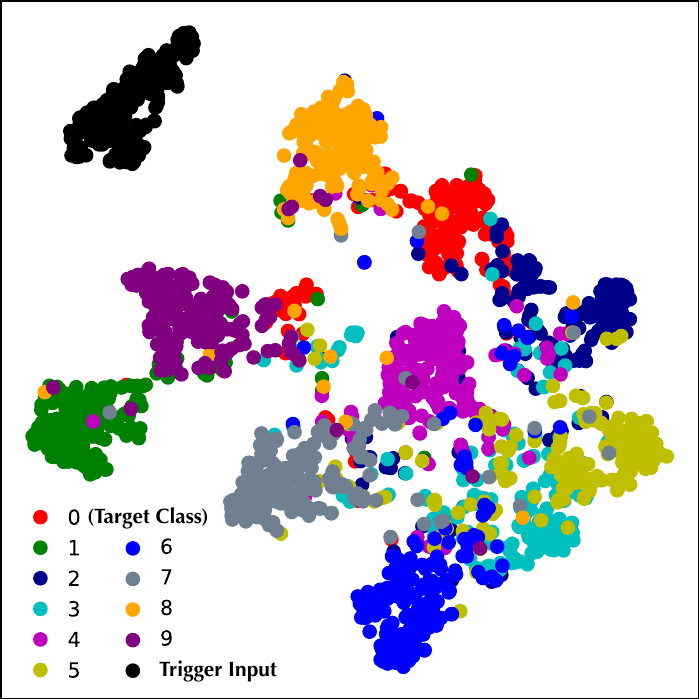}%
\caption{$t$-SNE visualization of the features of trigger-embedded and clean inputs in the supervised backdoor attack (target-class input: red; trigger-embedded input: black).}
\label{figure:sl-feature}
\end{figure}


To validate the analysis, we use \system to generate the poisoning data, pollute 1\% of the training data across all the classes, and train the model in a supervised setting for 20 epochs, which achieves 100\% \asr and 83\% \acc. We use $t$-SNE\mcite{van2008visualizing} to visualize the representations of trigger-embedded and clean inputs in the test set, with results shown in Figure\mref{figure:sl-feature}. Although the clusters of trigger inputs (black) and target-class inputs (red) are assigned the same label, they are well separated in the feature space, indicating that  supervised backdoor attacks do not necessarily associate the trigger with the target class in the feature space. This finding also corroborates prior work\mcite{tact}.

\begin{figure}[ht]
\centering
  \includegraphics[width=0.7\columnwidth]{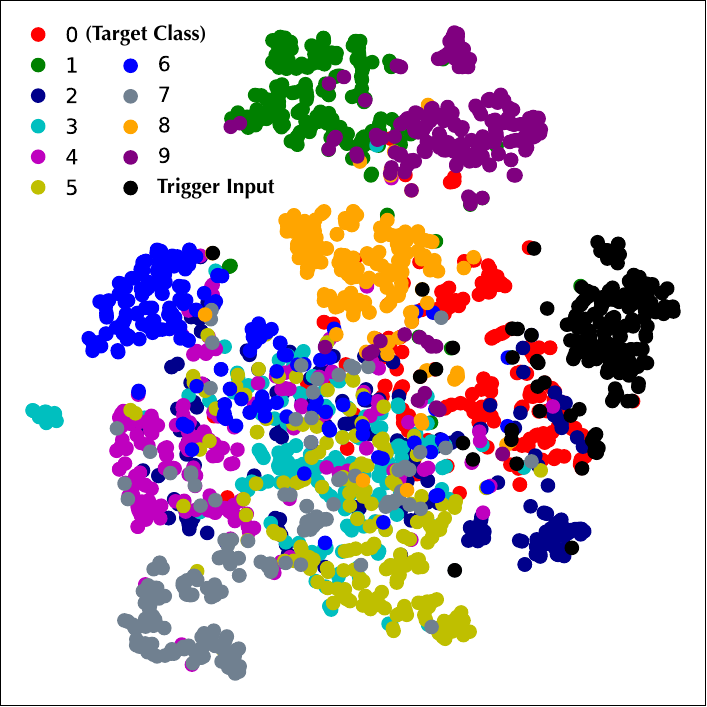}%
\caption{$t$-SNE visualization of the features of trigger-embedded and clean inputs in the self-supervised backdoor attack (target-class input: red; trigger-embedded input: black).}
\label{figure:ssl-feature}
\end{figure}

For comparison, we perform \system against SimCLR on CIFAR-10 and use $t$-SNE to visualize the features of trigger-embedded inputs and clean inputs in the test set, with results shown in Figure\ref{figure:ssl-feature}. Observed that in comparison with the supervised backdoor attack (\mcf  Figure\ref{figure:sl-feature}), the cluster of trigger-embedded inputs (black) and the cluster of target-class clean inputs (red) are highly entangled in the feature space, indicating that the self-supervised backdoor attack takes effect by aligning the representations of trigger-embedded inputs and the target class.

\section{Proofs}
\label{sec:proofs}

We first introduce the following two assumptions commonly observed in encoders trained using {\ssl}\mcite{alignment}

\begin{assumption}
\label{assump:align}
(Alignment) A well-trained encoder $f$ tends to map a positive pair to similar features. Formally, for given input $x$, $f(x)^\sinv f(x^\splus) \geq (1 -\epsilon)$, where $\epsilon \in [0, 1)$ is a small non-negative number. In particular, by design, the trigger $r$ is invariant to the augmentation operator: $f(r)^\sinv f(r^\splus) = 1$.
\end{assumption}

\begin{assumption}
\label{assump:concentration}
(Uniformity) A well-trained encoder $f$ tends to map inputs uniformly on the unit hyper-sphere $\gS^{d-1}$ of the feature space, preserving as much information of the data as possible. Thus, as the number of data points is large, the average angle $\theta$ between the features of a negative pair follows the distribution density function\mcite{concentration}: 
\begin{equation}
    h(\theta) = \frac{1}{\sqrt{\pi}} \frac{\Tau(\frac{d}{2})}{\Tau(\frac{d-1}{2})} (\sin \theta)^{d-2}, \quad \theta \in [0, \pi]
\end{equation}
As the dimension $d$ is high, most of the angles heavily concentrate around $\pi/2$.  
\end{assumption}


Based on \mref{assump:align} and \mref{assump:concentration}, we now prove Theorem\mref{the:attack_main}.




\label{proof}

\begin{proof}[Proof of Theorem\mref{the:attack_main}]
According to Assumption\mref{assump:align}, we have
\begin{equation}
f(x_\star)^\sinv f(x_\star^\splus) \geq (1 - \epsilon)
\end{equation}
In other words,
\begin{equation}
    \begin{aligned}
    &(1 - \alpha)^2f(x)^\sinv f(x^+) + \alpha^2 f(r)^\sinv f(r^+) + \\ \nonumber
    &\alpha (1 - \alpha) (f(x)^\sinv f(r^+) + f(r)^\sinv f(x^+)) \geq ( 1 - \epsilon). \\ \nonumber
    \end{aligned}
\end{equation}
Since both $f(x)^\sinv f(x^+)$ and $f(r)^\sinv f(r^+)$ are no larger than 1, we have
\begin{equation}
    \begin{aligned}
    &\alpha (1 - \alpha) (f(x)^\sinv f(r^+) + f(r)^\sinv f(x^+)) \\ 
    &\geq ( 1 - \epsilon) -  (1 - \alpha)^2f(x)^\sinv f(x^+)- \alpha^2 f(r)^\sinv f(r^+) \\ 
    &\geq ( 1 - \epsilon) - (1 - \alpha)^2 - \alpha^2 \\ 
    & = 2\alpha (1 - \alpha) - \epsilon \nonumber
    \end{aligned}
\end{equation}
Then, based on Assumption\mref{assump:align}, we have
\begin{equation}
    \begin{aligned}
    &f(x)^\sinv f(r) \geq 1 - \frac{\epsilon}{2\alpha(1 - \alpha)}.
    \end{aligned}
\label{eq:align}
\end{equation}
For $\tilde{x}_*$ and $x$, we have
\begin{equation}
    \begin{aligned}
    f(\tilde{x}_*)^\sinv f(x) &= (1 - \alpha)f(\tilde{x})^\sinv f(x) + \alpha f(r) ^\sinv f(x) \\ \nonumber
    \end{aligned}
\end{equation}
Since $\tilde{x}$ and $x$ are a negative pair, based on Assumption\mref{assump:concentration}, we have 
\begin{equation}
    \begin{aligned}
 \sE[f(\tilde{x}_*)^\sinv f(x)] &\geq \alpha (1 - \frac{\epsilon}{2\alpha(1 - \alpha)})\\
          & \geq \alpha -    \frac{\epsilon}{2(1 - \alpha)}
    \end{aligned}
    \label{eq:test}
\end{equation}

For a well-trained $f$, $\epsilon$ is a constant. Thus, both \meq{eq:align} and \meq{eq:test} are functions that first increase and then decrease with respect to $\alpha \in (0, 1)$.

\end{proof}

\section{Details of Experimental Setting}
\label{app:setting}
{\bf Dataset --} For each dataset, we split it as a training set and a testing set according to its default setting. Specifically, both CIFAR-10 and CIFAR-100 are split into 50,000 and 10,000 images for training and testing, respectively; ImageNet-100 is split as 130,000 training and 5,000 testing images; while GTSRB is split into 39,209 training and 12,630 testing images.

{\bf Data augmentation --} For convenience, we describe the details of data augmentations
in a PyTorch style. Specifically, following prior work\mcite{chen:2021:exploring, chen:2020:simple}, we use geometric augmentation operators including \textsl{RandomResizeCrop} (of scale [0.2, 1.0]) and \textsl{RandomHorizontalFlip}. Besides, we use \textsl{ColorJitter} with [brightness, contrast, saturation, hue] of strength [0.4., 0.4, 0.4, 0.1] with an application probability of 0.8 and \textsl{RandomGrayscale} with an application probability of 0.2. 

{\bf Encoder training --} We use the training set of each dataset to conduct contrastive learning. We show the hyper-parameters setting for each contrastive learning algorithm in Table\mref{tab:hyper_training}, which is fixed across all the datasets.

\begin{table}[!ht]{\small
\centering
\renewcommand{\arraystretch}{1.1}
\begin{tabular}{c|ccc}
\multirow{2}{*}{Hyper-parameter} & \multicolumn{3}{c}{SSL Method} \\
\cline{2-4}
                                & SimCLR  & BYOL   & SimSiam \\ \hline
Optimizer                       & SGD     & SGD    & SGD     \\
Learning Rate                   & 0.5     & 0.06   & 0.06    \\
Momentum                       & 0.9     & 0.9    & 0.9     \\
Weight Decay                    & 1e-4    & 1e-4   & 5e-4    \\
Epochs                          & 500     & 500    & 500     \\
Batch Size                      & 512     & 512    & 512     \\
Temperature                     & 0.5     & -      & -       \\
Moving Average                 & -       & 0.996  & -       \\ \hline
\end{tabular}
\caption{Hyper-parameters of encoder training. \label{tab:hyper_training} }}
\end{table}

{\bf Classifier training --} Without explicit specification, we randomly sample 50 examples  from each class of the corresponding testing set to train the downstream classifier. We show the hyper-parameters of classifier in Table\mref{tab:hyper_classifier}.

\begin{table}[!ht]{\small
\renewcommand{\arraystretch}{1.1}
\centering
\begin{tabular}{c|c}
Hyper-parameter & Setting            \\ \hline
Optimizer      & SGD              \\
Batch Size     & 512              \\
Learning Rate  & 0.2              \\
Momentum       & 0.9              \\
Scheduler      & Cosine Annealing \\
Epochs         & 20               \\ \hline
\end{tabular}
\caption{Hyper-parameters of classifier training. \label{tab:hyper_classifier} }}
\end{table}

{\bf Evaluation --} We evaluate ACC using the full testing set. For ASR, we apply \system on the full testing set and measure the ratio of trigger inputs that are classified to the target class.
All the experiments are performed on a workstation equipped with Intel(R) Xeon(R) Silver 4314 CPU @ 2.40GHz, 512GB RAM, and four NVIDIA A6000 GPUs.


\section{More Experimental Results} 
\label{app:details}

Here, we show the additional experimental results.

{\bf Performance of clean models --} The ACC and ASR of clean models trained by SimCLR, BYOL, and Simsiam are summarized in Table\mref{clean_acc}.

\begin{table}[!ht]{\small 
\centering
\renewcommand{\arraystretch}{1.1}
\setlength{\tabcolsep}{2pt}
\begin{tabular}{c|cccccc}
\multirow{3}{*}{Dataset} & \multicolumn{6}{c}{SSL Method}                                                      \\ \cline{2-7} 
                         & \multicolumn{2}{c}{SimCLR} & \multicolumn{2}{c}{BYOL} & \multicolumn{2}{c}{SimSiam} \\ \cline{2-7} 
                         & ACC           & ASR        & ACC          & ASR       & ACC            & ASR        \\ \hline
CIFAR-10                 & 79.1\%        &   9.93\%         & 82.4\%       &   12.2\%        & 81.5\%         &   11.75\%         \\
CIFAR-100                & 48.1\%        &    1.14\%        & 51.0\%       &   0.46\%        & 52.0\%         &   0.72\%         \\
ImageNet-100             & 42.2\%        &    1.59\%       & 45.1\%       &     1.41\%      & 41.3\%         &   1.53\%        
\end{tabular}
\caption{
Accuracy of different SSL methods under normal training. 
}
\label{clean_acc}}
\end{table}

\begin{figure}[!t]
\centering
  \includegraphics[width=1\columnwidth]{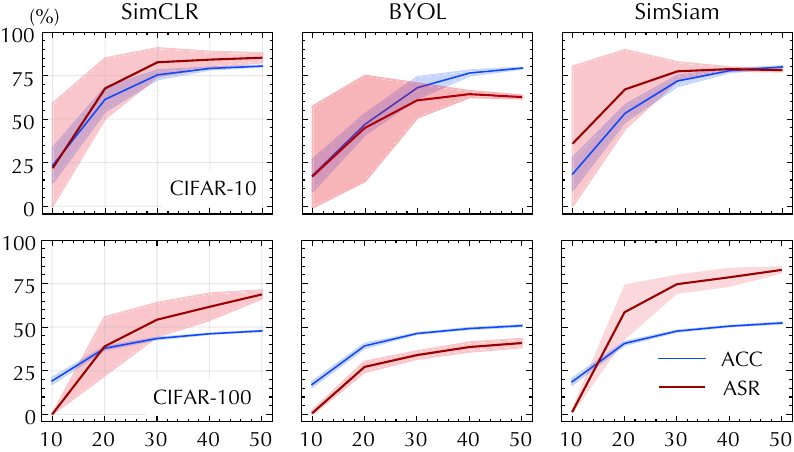}%
\caption{Performance of \system w.r.t. the fine-tuning data size. \label{figure:number_labeled}}
\end{figure}

{\bf Fine-tuning data size --} Typically, equipped with the pre-trained encoder, the victim fine-tunes the downstream classifier with a small labeled dataset. Here, we evaluate the impact of this fine-tuning dataset on \system. Figure\mref{figure:number_labeled} illustrates the performance of \system as a function of the number of labeled samples per class.


Observe that both \acc and \asr of \system increase with the fine-tuning data size, while their variance decreases gradually. For instance, when the number of labeled samples per class is set as 50, the ASR of \system on 
CIFAR-10 under SimSiam stably remains around 75\%. This may be explained as follows. Without the supervisory signal of labeling, \system achieves effective attacks by entangling the representations of trigger-embedded and target-class inputs (details in \msec{sec:explore}). During fine-tuning, more labeled samples imply that the representations of trigger-embedded inputs are more likely to be associated with the target-class label, leading to higher and more stable \asr. In other words, more fine-tuning data not only improves the model's performance but also increases its attack vulnerability.

\begin{figure}[!ht]
	\centerline{\includegraphics[width=1\columnwidth]{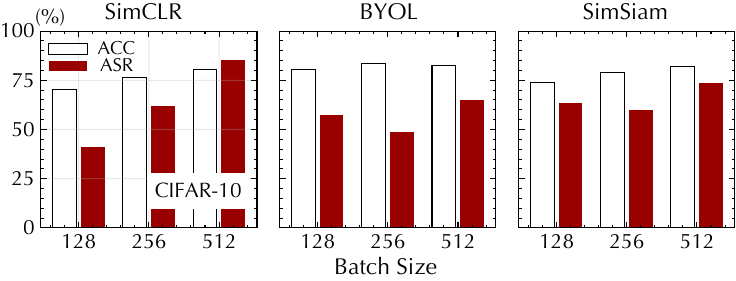}}
	\caption{Performance of \system w.r.t. the batch size on CIFAR10.}	
	\label{figure:batchsize}
\end{figure}

{\bf Batch size --} Existing studies show that batch size tends to impact the performance of contrastive learning\mcite{chen:2021:exploring}. Here, we explore its influence on the performance of \system. Specifically, on the CIFAR-10, we measure the \acc and \asr of \system with the batch size varying from 128 to 512, with results shown in Figure\mref{figure:batchsize}.

Observe that the model's accuracy improves with the batch size, which corroborates the existing studies\mcite{chen:2021:exploring}. Moreover, a larger batch size (\meg, $\geq 512$) generally benefits the \asr of \system. This may be explained by that more positive pairs (also more negative pairs in SimCLR) in the same batch lead to tighter entanglement between trigger-embedded and target-class inputs. Meanwhile, for smaller batch sizes (\meg, $\leq 256$), the three \ssl methods show slightly different trends. This may be attributed to the design of their loss functions: BOYL and SimSiam only optimize positive pairs, while SimCLR optimizes both positive and negative pairs, thereby gaining more benefits from larger batch sizes.

\begin{figure}[!ht]
\centering
 \includegraphics[width=1\columnwidth]{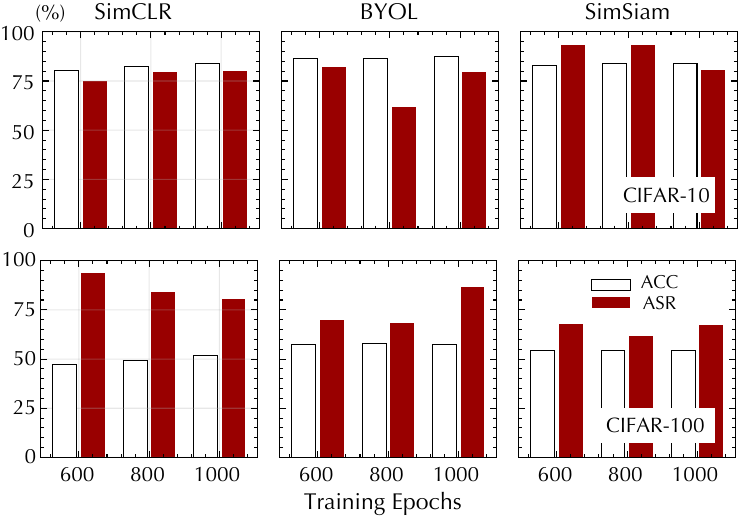}%
\caption{Performance of \system w.r.t. the number of epochs.}
\label{figure:acc_epoch}
\end{figure}

{\bf Training epochs --}
Typically, SSL benefits from more training epochs\mcite{chen:2020:simple}. We evaluate the impact of training epochs on the performance of \system. Figure\mref{figure:acc_epoch} shows the ACC and ASR of \system as the number of epochs varies from 600 to 1,000.

Observe that as the training epoch increases, the ACC of \system gradually grows, while the ASR remains at a high level. For example, on CIFAR-10 with SimCLR, when the number of epochs increases from 600 to 1,000, the ACC also increases from 80.52\% to 83,94\%, and the ASR remains above 75\%. In a few cases, the ASR slightly drops. We speculate this is caused by the random data augmentations used in \ssl. Side evidence is that on CIFAR-10 with BYOL, the ASR first slightly decreases and then remains above 80\%. In general, the number of training epochs has a limited impact on the effectiveness of \system.

\end{document}